\input epsf

\documentstyle[12pt]{article}

\advance \voffset by -1.5 in
\advance \hoffset by -1 in

\font\smalli=cmr8 scaled\magstep1
\def\thecaption#1#2{\centerline{\vbox to 1 in{\hsize 5 in \vfill
           {\textindent{#1} \global \advance \baselineskip by -10 pt
          \smalli \noindent#2 }}}\global \advance \baselineskip by 10 pt}

        \newcommand{\be}{\begin{equation}}
        \newcommand{\ee}{\end{equation}}
        \newcommand{\ba}{\begin{eqnarray}}
        \newcommand{\ea}{\end{eqnarray}}
        \newcommand{\ban}{\begin{eqnarray*}}
        \newcommand{\ean}{\end{eqnarray*}}
        \newcommand{\Ba}{\begin{equation} \begin{array}}
        \newcommand{\Ea}{\end{array} \end{equation}}
\def\lesim{\,{\raise-3pt\hbox{$\sim$}}\!\!\!\!\!{\raise2pt\hbox{$<$}}\,}

\def\d{\partial}

\def\a{\alpha}
\def\b{\beta}
\def\g{\gamma}

\def\e{\epsilon}
\def\et{\eta}
\def\t{\theta}
\def\vt{\vartheta}
\def\k{k}

\def\la{\lambda}
\def\j{j}
\def\m{\mu}
\def\n{\nu}
\def\r{\rho}
\def\s{\sigma}

\def\c{\chi}
\def\p{\psi}
\def\o{\omega}
\def\L{\Lambda}
\def\cO{\cal O}
\def\Th{\Theta}
\def\Om{\Omega}

\def\Vc{{\cal V}}
\def\ps{\not \! p}
\def\Js{\not \! J}

\def\Us{\not \! \Upsilon}
\def\pp{{\mbox{\raisebox{-.05cm}{$\stackrel{\perp}{p}$}}}}
\def\ppm{{\mbox{\raisebox{-.05cm}{$\stackrel{\perp}{p}$}}}
         {\mbox{\raisebox{.01cm}{$^{\;\mu}$}}}}
\def\ppa{{\mbox{\raisebox{-.05cm}{$\stackrel{\perp}{p}$}}}
         {\mbox{\raisebox{.01cm}{$^{\; a}$}}}}
\def\ppb{{\mbox{\raisebox{-.05cm}{$\stackrel{\perp}{p}$}}}
         {\mbox{\raisebox{.01cm }{$^{\; b}$}}}}

\def\ppdo{{\mbox{\raisebox{-.05cm}{$\stackrel{\perp}{p}$}}}
         {\mbox{\raisebox{.01cm}{$^{\; 2}$}}}}

\def\pps{\not \! {\mbox{\raisebox{-.05cm}{$\stackrel{\perp}{p}$}}}}
\def\ls{\not \! \lambda}
\def\ov{\over}
\def\nn{\nonumber}

\begin{document}

%
%

	\begin{center}
	{\LARGE Neutrino Oscillations \\
		in \\
		Strong Gravitational Fields \\}
	\vspace{1.3cm}
	{\large Dardo  P\'{\i}riz, Mou Roy and Jos\'{e} Wudka \\}
	\vspace{.4cm}
	{\small {\it Department of Physics \\
	University of California, Riverside \\
	California 92521-0413, U.\ S.\ A.  \\ 
	March 14, 1996.\\}}
	\vspace{1.8cm}
	\end{center}

\begin{abstract}

Neutrino oscillations in the presence of strong gravitational fields are
studied. We look at very high energy neutrinos ($\sim $1 TeV) emanating from
Active Galactic Nuclei (AGN). It is observed that spin flavor
resonant transitions of such neutrinos may occur in the vicinity
of AGN due to {\it gravitational} effects and due to the presence of a
large magnetic field ($\sim $1 Tesla). A point to note is that matter
effects (normal MSW transitions) become negligible in comparison
to gravitational effects in our scenario.

\end{abstract}

\newpage
\global\advance\baselineskip by -5 pt

\begin{center}
\subsection*{1. Introduction}
\end{center}

It is well known that relativistic effects produce a precession of
gyroscopes in the vicinity of a source of gravitational field \cite{gyro}.
It is therefore reasonable to expect that particles will undergo
helicity flips in such regions. This is particularly interesting in
the case of
neutrinos because left handed neutrinos traversing a
strong gravitational field could be converted into unobservable
right handed ones, resulting in a
decrease of the neutrino flux\footnote{Left-right
 transitions can occur as a gravitationally induced coherent precession, as
 was investigated by Cai and Papini, \cite{cp}, and is in general quite
small.}.
We will analyze in this paper, following \cite{wud},
the possibility of encountering resonant transitions  of
high energy Dirac neutrinos produced in  Active Galactic Nuclei (AGN),
(reminiscent of the ones proposed as solutions of the solar neutrino problem
\cite{{lw},{bb},{ms}}). The presence of such resonances alters the
expectation of the neutrino flux on Earth. In our calculations we will
consider magnetic, matter and gravitational effects. The latter are
introduced using the standard couplings of neutrinos to gravity; we will
not consider possible effects of the modification of the equivalence
principle \cite{equiv.prin.} in this paper.

The search for such high energy neutrinos by the neutrino telescopes
(e.g. DUMAND, AMANDA, NESTOR, BAIKAL, etc., \cite{jw})
which are under construction necessitates a clear picture of the expected
neutrino fluxes from these objects.
AGN are by far the strongest sources of ultrahigh
energy neutrinos in the Universe, producing fluxes
expected to be detectable on the earth
with present technology \cite{jw}. These objects are believed to be
fueled by the gravitational energy of matter accreting onto a supermassive
black hole ($10^4$ to $10^{10} M_{\odot}$) at the AGN core, where
gravitational energy is converted into luminous energy through the
acceleration of high energy protons \cite{zp}. This environment exhibits
copious production of hadronic and subsequent leptonic high energy
byproducts, such as neutrinos.

The paper is organized as follows. We start with a brief discussion of
an AGN model and mechanism of high energy neutrino production in it
(section 2).
In section 3 we study the possibility of neutrino transitions between
different flavors and spins in the environment of a strong gravitational
field using the semiclassical approximation. In section 4 we apply the
formalism to the interesting case of a Kerr black hole;
our main results are presented in this
section. Finally in section 5 we give our conclusions. Several
mathematical details are relegated to the appendices.

\begin{center}
\subsection*{2. AGN Model}
\end{center}

Active galactic nuclei have long been recognized as possible sources
of high energy signals \cite{man},
 being the most luminous objects in the Universe.
They have luminosities ranging from $10^{42}$
to $10^{48}$ ergs$/$sec, corresponding to black hole masses of the order
of $10^4 $ to $10^{10} M_{\odot}$, on the natural assumption that they are
powered by Eddington-limited accretion onto the black hole.
The spherical accretion model
(based on  works by Kazanas, Protheroe and Ellison
\cite{{pk},{ke}}) is used in most of the calculations
of the neutrino production in central regions of AGN \cite{{zp},{br},{sb}}.
 According to this scenario, close to the black hole the accretion flow becomes
spherical and a shock is formed where the ram pressure of the accretion flow
is balanced by the radiation pressure.
We will follow this model
even though it is only approximately true considering
that we will be looking at rotating black holes.

The distance from the AGN center to the shock, denoted as the shock
radius $ {\cal R}$ ($\simeq$ a few Schwarzschild radii)
contains the central engine of AGN.  The shock radius is parametrized
\cite{{pk},{ke}}
by $ {\cal R} = x  r_g$ where $ r_g $ is the Schwarzschild radius
of the black
 hole, and $x$  is estimated to be in the range of $10$ to $100$ \cite{zp}
(which is consistent with the available data \cite{wy}).

The matter density at the shock $\rho({\cal R})$ can be estimated
from the accretion
rate needed to support black hole luminosity, and from the radius and
accretion velocity at the shock \cite{zp}
\be \rho \left( {\cal R} \right)
 \simeq 1.4 \times 10^{33} x^{-{5 \over 2}} L^{-1}_{\rm AGN} Q^{-1}
\;{\rm gm/cm}^3   \label{dena} \ee
where $Q(x) = 1 - 0.1 x^{0.31}$ is the efficiency for converting
accretion power into accelerated particles at the shock \cite{ke},
and $L_{\rm AGN}$ is the AGN continuum luminosity in units of ergs/sec.
 In this model the matter density falls off as
\be \rho_1(r) = \rho\left( {\cal R} \right) \;
\left( { r \over {\cal R}} \right )^{ - {3 \over 2}}; \quad r > {\cal R }.
\label{denb}    \ee

Acceleration of protons is assumed to occur by the first order diffusive Fermi
mechanism at the shock \cite{ke}. Energy losses
during acceleration occur through {\it pp} collisions in the gas
\footnote{Accelerated protons interact with
protons of the accreting plasma to form charged pions.}; and also
through the $ p \gamma \rightarrow p + e^+ + e^- $
and $ p \gamma \rightarrow N \pi $ processes
in the dense radiation fields present, mainly, in
the central region. All these reactions give
rise to high energy neutrinos through the $ \pi ^\pm \rightarrow \mu^\pm
\rightarrow e^\pm$ decay chain \cite{zp,pk,e,s}.
These neutrinos are expected to dominate the neutrino sky
\cite{zp} at energies of 1 TeV and beyond.

Since neutrons are not confined by the magnetic field,
they are free to escape
from the central core  \cite{zp}. But neutrons
can also produce high energy neutrinos
{\it via\/} $ np $ and $n \gamma $ collisions.
The conclusion of ref. \cite{zp} was that
neutrons with energy less than approximately
$10^3$ TeV do escape the central core. On the other hand, Stecker
{\it et. al.} \cite{sd}, found that secondary neutrons will not in general
escape the shock radius, and argue that
a significant amount of power is generated through
$ n \gamma $ interactions. However, according to them, confinement of
nucleons
in the central core occurs only for energies beyond $\sim 10^4$ TeV.
In our study, as we shall see later, the matter density is irrelevant whatever
the status of neutrons, confined or not.

Magnetic effects around AGN have important astrophysical consequences.
To estimate a value of the magnetic field of AGN which we will use
in our calculations, we look at two specific models.
Both these models assume the ``equipartition" condition that the
external pressure is matched by the magnetic one.
According to Begelman {\it et.al.} \cite{bbr}, a characteristic  magnetic field $B$
is given by

\be
B = { 10^8  \ov \left({ M/M_{\odot}}\right)^{1/2}}\; {\rm G}
\label{mago}
\ee
where $M$ is the mass of the black hole considered.
Szabo and Protheroe \cite{zp} estimate the magnetic field at a shock
formed by accretion to be
\be
B \simeq 5.5 \times 10^{27} Q^{-{ 1 \ov 2 }} x^{-{ 7 \ov 4}}
L^{-{ 1 \ov 2}}_{\rm AGN} \; {\rm G}  \label{magb}
\ee
where the symbols have the same meaning, as in equation (\ref{dena}).
Since the shock is the site of high energy neutrino production
a good estimate of the value of magnetic field would be that
evaluated at the shock. In the absence of a detailed model we will
assume that the magnetic field remains approximately constant inside 
a region of size equal to the pressure scale height, the average fields
of two of these regions will be uncorrelated. 
For  $ L_{\rm AGN}=10^{45}$ ergs/sec,
$ M=10^{8} M_{\odot} $ and $x=10 $ both of the above expressions give 
$ B \sim 10^4 $ G. We will be interested in the vicinity of the
horizon where the pressure scale height is $ \sim r_g $. In this region
we will assume that $ B $ remains at the above value. Outside the
inner core of the AGN the magnetic field will drop as $({\rm pressure})^{1/2}$.

\begin{center}
\subsection*{3. Dirac Equation in Curved Space-Time and Neutrino Oscillations}
\end{center}

We will study the possibility of neutrino oscillations between different
flavors and spins in the environment of a strong gravitational field.
 In this paper we consider, for simplicity, the case of two family flavors
only. We will also restrict ourselves to the case of Dirac neutrinos
interacting minimally (as described by the equivalence principle) with
the gravitational field.

The outline of our method to compute neutrino oscillations
is the following. Starting  with
the Dirac equation in the presence of a strong gravitational field,
we use a semiclassical expansion to different orders in the relevant parameters
to determine the effective Hamiltonian for positive momentum states.
We look in particular for the lowest order off-diagonal terms
that could possibly induce gravitational neutrino oscillations,
including the effects of magnetic fields associated with AGN.
Once  the Hamiltonian is obtained, we analyze the possibility of resonances
depending on the neutrino mass, energy, angular momentum and the black hole
mass and angular momentum. The
different neutrino transitions and expected neutrino fluxes are then
investigated.

\begin{center}
\subsubsection*{3.1  Semiclassical Approximation}
\end{center}

We start with the Dirac equation in curved space-time~\cite{DW},
\be     [i e^\m_a \g^a (\d_\m + \o_\m) - m + \Js P_L ] \p = 0  \label{aaaa} \ee
where $e^\m_a$ are the tetrads, $m$ is the mass matrix, $ \Js $ is the
weak interaction current matrix,
 $P_L$ the left-handed projection operator, and
the spin connection is
\be
\o_\m={1\ov 8}[\g_a,\g_b] e^{\n a} e^b_{\n;\m}   \label{spinco}
\ee
where the semicolon denotes a covariant derivative;
we use Greek indices for coordinates
in the general frame and Roman indices for the local inertial frame.
As in flat space, there are four neutrino states for each flavor that we
 separate into two  states corresponding to neutrinos traveling towards
or away from the observer. We call these
{\it positive} and {\it negative} momentum states respectively.

The first step in the semiclassical approximation is to replace the neutrino
spinor by
\be
\psi = e^{i S} \chi \label{defofchi} \ee
where $S$ is the classical action defined in terms of
the Lagrangian
density ${\cal L}(x,\dot{x})$ and affine parameter {\it l} as
\ba
& S & = \int dl\,{\cal L}(x,\dot{x}) \hspace{1cm}
{\cal L} = - {1 \over 2} g_{\mu \nu} \dot{x}^\mu \dot{x}^\nu  ; \quad
\dot{x} = {dx \over dl}
\ea
(an over-dot will always denote differentiation with respect to $ l $),
and solves the Hamilton-Jacobi equation
\be g^{\mu \nu} \partial_\mu S \partial_\nu S = 0. \label{HJeq}\ee
$\chi$ is constructed as the solution of the Hamilton-Jacobi equation
and determines the trajectory of classical particles.

In order to proceed further we use the approach
described by Sakita and Tsani
\cite{st}.
If $ \bar{x} $ be the solution to the classical geodesic
equation for massless particles, that is the classical trajectory,
a new set of local coordinates
\be \{ l, \xi^A \},\; A = 1,2,3 \ee
is  chosen such that
\be x^\m = \bar{x}^\m(l) + \n^\m_A(l)\xi^A.  \ee
Here $\n^\m_A(l)$ solves the variation equation
\be {\partial_A\left\{\left({\d{\cal L}\ov \d x^\mu }\right)-
{d \ov dl} \left( { \d {\cal L} \ov \d \dot{x}^\mu } \right ) \right \}}= 0
\ee
to $O(\xi^2)$.
The functions $\bar x + \nu_A \xi^A $ solve $ \delta S = 0 $ to
second order in  $\xi$ and so describe (approximately)
a bundle of neighboring geodesics parametrized by the $ \xi^A $;
 the values of the $ \xi $ determine the
separation of a given geodesic in the bundle from the reference
geodesic $ \bar x^\mu( l ) $.

The semiclassical momentum is defined as
\be p_\mu = -\partial_\mu S \vert_{ x = \bar{x} } \ee
where the derivative is with respect to the end-point. The solution to
(\ref{HJeq}) is, to $ O ( \xi^2 ) $,
\be S= -p_\m \n^\m_A\xi^A-
{1\ov 2}(\bar{\Gamma}^\m_{\n\r}p_\m \n^\n_A \n^\r_B +
{1 \ov 2} \dot {N}_{AB} ) \xi^A \xi^B \ee
where $\bar{\Gamma}^\m_{\n\r}$ are the Christoffel symbols evaluated at
$x=\bar{x}$, and
\be N_{AB}= \nu^\mu_A \nu^\nu_B g_{\mu\nu}. \ee

From the geodesic equation obeyed by $p^\m $ it follows that
(see Appendix A)
\be p_\mu \nu^\mu_A = c_A = {\rm constant}. \label{pnu}  \ee
In the following it will prove convenient to define a time-like
vector $\ppm$ which is the component of
momentum $p$ orthogonal to the $ \n^\m_A$, namely
\be \ppm = p^\mu - c_A ( N^{-1})^{AB} \nu^\mu_B. \ee

\begin{center}
\subsubsection*{3.2 Effective Hamiltonian}
\end{center}

With the above preliminaries we are in a position to evaluate the
effective Hamiltonian for the neutrino system. In order to make a
systematic expansion let us define {\it R} to be the scale of the metric,
so that for example, $ \omega_\m \sim { 1/R}$; and let {\it p} be the
order of magnitude of the momentum of the neutrinos. We imagine a localized
wave packet of extension $\xi$ propagating freely through a region of
size {\it R}, where the gravitational field is essentially constant.
The uncertainty relation requires that a change in the momentum is related
to {\it R via} $\Delta p \sim 1/R$. This in turn implies that the angular
spread of the wave packet is given by $\Delta \t = \Delta p /p \sim 1/(pR)$,
so that in a distance {\it R} a wave packet spreads $R \Delta \t \sim 1/p$.
But this spatial spread is determined by the values of $\xi$ for which the
states are significantly different from zero, so that $\xi \sim 
1/p$~\cite{anandan}.

Going back to the Dirac equation in curved space (\ref{aaaa}),
after substituting (\ref{defofchi}), we obtain the equation
\be ( - e^{\m}_a \g^a \partial_{\m} S + i e^{\m}_a \g^a \partial_{\m}
+ \Vc_0 - m   )\chi = 0 \ee
where
\be \Vc_0 = i \g^a e^\m_a \o_\m + \Js P_L \label{pote} \ee
and $ \omega_\mu $ is defined in (\ref{spinco}).

We make a double expansion of $\c$, first in powers of
$\xi$ and then in powers of $1/pR$,
as follows

\Ba{rclll}
\vspace{0.2cm}
\chi&=&\chi^{(0)}+\chi^{(1)}_A \xi^A + O \left( \xi^2 \right) & &  \nn \\
\chi^{(0)}&=&U_0+U_{1 \ov 2}+ U_1 + \cdots ; &
 U_\n \sim {(pR)^{-\n}} & \n = 0, {1\over2} , 1 ,\cdots  \nn \\
\chi^{(1)}_A&=&V_{1A}+V_{{ 3 \ov 2}A} + V_{2A} + \cdots ; &
V_{\n A} \sim { R^{-1} (pR)^{1-\n}} & \n = 1 , {3\over2} , 2 , \cdots.
\label{expa}
\Ea

We substitute these expressions into the Dirac equation and make a
double Taylor expansion in $ \xi $ and $ 1 / ( p R ) $ and demand
that each term vanish separately.
In the perturbation we choose the mass to be $m \sim (p/R)^{1/2}$.
This is for convenience in defining the perturbation expansion
in an effort to look for the lowest order off-diagonal terms
that might cause gravitationally induced helicity flips of neutrinos.
For the lowest possible magnitude of the mass, {\it i.e.} $m\sim 1/R$, 
there are  no off diagonal terms in the Hamiltonian.

The result is the following set of mixed equations for the spinors
$ U_0, U_{1\ov 2}\cdots$.
To $O(\xi ^0)$,
\ba
&&\ps U_0 = 0,  \qquad  \ps U_{ 1 \ov 2}= m U_0, \nn \\
&&\ps U_1 + { i \ov \pps}\dot{U_0} + i\Us^A V_{1A} + \bar{\Vc}_0 U_0 -
mU_{1\ov 2} = 0  \nn	\\
&& \ps U_{ 3 \ov 2} + { 1 \ov \pps} U_{1 \ov 2}+ i\Us^A V_{{ 3 \ov 2} A}
+ \bar{\Vc}_0 U_{ 1 \ov 2} - m U_1  = 0      \label{aro}
\ea
where
\be
\Us^A = (N^{-1})^{AB} (\nu_{\m B} - { c_B \pp_{\m} \ov {\ppdo}})
\bar{e}^{\m}_a \g^a.
\label{poteo}
\ee
(the over-bar represents variables evaluated on the geodesic:
$ x = \bar x $).
To $O(\xi)$,
\be
\ps V_{1A} = \ls_A U_0,  \qquad
\ps V_{ {3 \ov 2} A} = \ls_A U_{1\ov 2} \label{deb}
\ee
where
\be
\ls_A = { 1 \ov 2} \dot{N}_{AB}\Us^B - \bar{e}^{\m}_{a;\n}
p_{\m} \nu^{\n}_A\g^a .
\ee
From  (\ref{aro}) and (\ref{deb}) we obtain
\ba
& i \dot{U}_0 & = {\cO } U_0 	\nn	\\
& i \dot{U}_{ 1 \ov 2} & = {\cO } U_{1 \ov 2} - { im \ov 2} \Us^A V_{1A}
- { m \ov 2} \bar{\Vc}_0 U_0.	\label{red1}
\ea
where
\be
{\cO } = -{ 1 \ov 2} \ps \bar{\Vc}_0 + { m^2 \ov 2} - { i \ov {2 \pps}}
\dot{\ps} + {i  \ov 2}\Us^A\ls_A \label{red2}
\ee
It is now possible  to reduce these equations to a
Schr\"{o}dinger-like equation (as shown in Appendix B)
involving only $\chi$ which reads
\be i \dot\chi =\left(  {\cO } - { m \over 2} \bar{\Vc}_0 \right)
\chi.    \label{schr}    \ee
The above equation describes a set of states which are almost pairwise
degenerate (due to the condition $ \ps \chi \simeq 0 $); the effective
Hamiltonian $ \tilde{H}_{\rm eff} $
for the positive momentum states is obtained using degenerate
perturbation theory. Denoting by $P_+$ the projector onto the positive
momentum states we find
\be
\tilde{H}_{\rm eff}  P_+ =  P_+ \left(  {\cO } - { m \over 2} \bar{\Vc}_0 
\right) P_+  . \label{eff}
\ee
It is always possible to go to the $\pp$ rest frame where
$\pp{}^a =E(1,0,0,0)$
and $p^a =E(1,0,0,1)$, neglecting the neutrino rest mass
and using
$\ppdo  = p\cdot \pp$; in this (local Lorentz) frame,
the projector onto positive
momentum states is
\be P_+ = diag(1,0,0,1) \label{pplusdef} \ee
In what follows we choose frames where the momentum
has the form given above. In Appendix C we show that, with these
previous considerations, equations (\ref{schr}) and (\ref{eff}) lead to
\be \tilde{H}_{\rm eff} = i \dot{\alpha} + { 1 \ov 2} m^2 - p \cdot J_{\rm eff} P_L + m \Th^b \tau_b 	\label{heff}	\ee
where $\dot{\alpha}$ is spin and flavor diagonal and has no observable
consequences~\footnote{An imaginary contribution to $ \alpha $ would
signal a decrease in the overall neutrino flux and is also
unobservable.}.
The effective current is
\be J_{\rm eff}^a = \bar{J}_W^a - \bar{J}_G^a.  \label{jefa}   \ee
Here $\bar{J}_W^a$ is the weak interaction current, $\bar{J}_G^a$ is defined
as
\be J^a_G  = { 1 \ov 4} \e^{abcd} \lambda_{fcd} \left( \eta^f_b +
{ 2 p^f \pp_b \ov \ppdo} \right )      \label{jefb}  \ee
and
\be
\lambda_{fcd} = ( e_{f\m ,\n} - e_{f\n,\m}) e^{\m}_c e^{\n}_d,\;\;\;
 \eta^{ab} = diag(1,-1,-1,-1) \label{you}
\ee

In order to determine $\bar{J}_W^a $ we
note that in the {\it rest frame of the accreting matter} it takes the form
$ J^0_{W; \nu_e}\;=\;  G_F ( 2 N_p - N_n) / \sqrt{2}   $
and $ J^0_{W; \nu_\mu}\;=\; - G_F N_n /  \sqrt{2} $ where $G_F$ is the
Fermi coupling constant and $N_p$ and $N_n$ are the
proton and neutron number densities respectively (only the
zero-th component is non-vanishing in this frame). In the neutrino frame
used above this becomes
 \be J^a_{W; \nu_e}\;=\; { G_F \ov \sqrt 2} ( 2 N_p - N_n) \; u^a ,
\qquad J^a_{W; \nu_\mu}\;=\; - { G_F \ov \sqrt 2} N_n \; u^a \label{den} \ee
where  $u^a$ denotes the four velocity of matter in the chosen local
inertial frame. In the model considered, matter is ultrarelativistic \cite{pk}
(within the region of interest), so that $ u^2 \simeq 0 $.

Finally,
\be \Th^b = { 1 \ov 2}\bar{e}^{\m}_{a;\m} \e^{acd} { p_c \pp_d \ov \ps^2}
+ { 1 \ov 2} \left( \bar{J}^a_W +  { 1 \ov 2} \e^{acde}\bar{ \g}_{cde}\right )
\left(\eta^b_
a + {p_a p^b - \pp_a p^b - p_a \ppb  \ov \ppdo }\right ).  \label{jefc}    \ee

In addition to the gravitational effects it is essential to incorporate
the effects of the magnetic field associated with AGN.
It is well known that if a neutrino has non-vanishing magnetic moment (or transition magnetic moment of dipole type)
$\mu_\nu$\footnote{Okun \cite{okun} pointed out that the manifestations of an
electric dipole moment and magnetic dipole moment are practically
indistinguishable for a highly relativistic neutrino. For simplicity
the term ``magnetic moment'' will be used in this paper to represent
the combined effect of an electric and magnetic dipole moment.},
its helicity can be flipped when it passes through a region with a
magnetic field which has a component perpendicular to the direction
of motion. This idea has been analyzed by Okun {\it et.al.} \cite{okun},
and more recently the combined effects of flavor mixing, magnetic spin flip
and matter interactions have been considered, \cite{lm,ak}.
The interaction
with the electromagnetic field stems from a term of the form
\be  \mu \sigma^{ a b } F_{ a b } \psi \label{EMterm} \ee
($ F_{ a b }$ is the electromagnetic field tensor
and $ \sigma^{ a b }= {1\ov 4}[\g^a,\g^b]$.) to be added to the left-hand
side of (\ref{aaaa}); just as in the above references only the electric
and magnetic fields in the direction orthogonal to $ \vec p $
contribute. Denoting by $E^r$ and $B^r$
the components of the electric and magnetic fields respectively
in the frame where $ p^a = E ( 1 , 0 , 0 , 1 ) $, and using
(\ref{EMterm})  we find that the
effective Hamiltonian is modified according to 
\be 
\tilde{H}_{\rm eff} \rightarrow \tilde{H}_{\rm eff} + 
\mu \sqrt{ \pp \cdot p } \, \left( \begin{array}{cc}
0 & \Omega^* \\ \Omega & 0 
\end{array} \right) ; \qquad
\Om = \left( B^1 - E^2 \right) + i \left( E^1 + B^2 \right)
\label{defofOm} \ee
where $ \mu $ denotes the magnetic moment matrix and $ B^a , \, E^a $
denote the magnetic and electric fields measured by a locally inertial
observer.

We consider two generations with four-component Dirac neutinos.
Here we examine the  $ \nu_e-\nu_\mu $ system. The same can be
done for the $ \nu_e-\nu_\tau  $ and $ \nu_\mu-\nu_\tau  $ systems.
Using the chiral bases  $\nu_{e_L}, \nu_{\mu_L}, \nu_{e_R}, \nu_{\mu_R} $,
we can  write the evolution equation for neutrino propagation
through matter in the presence of a strong gravitational field as \footnote{
Note that the left hand side involves differentiation with respect to the
affine parameter which has units of $ E^{-2} $ the units of length being
$ E^{-1} $ . Therefore $\tilde{H}_{\rm eff}$ has units of $ E^{2} $ which
differs from the usual units of the Hamiltonian.}
\be
i {d \ov dl} \left( \begin{array}{c}
 \nu_{e_L}\\ \nu_{\mu_L}\\  \nu_{e_R}\\  \nu_{\mu_R}
\end{array}
\right ) = \tilde{H}_{\rm eff} \left( \begin{array}{c}
\nu_{e_L}\\ \nu_{\mu_L}\\  \nu_{e_R}\\  \nu_{\mu_R}
\end{array}\right ) 
\ee
where $\tilde{H}_{\rm eff}$ is  the $4 \times 4$ matrix (\ref{heff}), 
containing the effects of
the weak and gravitational interactions and the effects of
electromagnetic field.
An order of magnitude estimate (supported by explicit computations for
particular metrics) show that the $\Th $ terms (eq. \ref{jefc}, off-diagonal
terms due to gravitational effects) in the effective Hamiltonian are
negligible compared to the effects of the magnetic field
for all interesting cases (see section 3.4). The gravitational effects 
in the diagonal elements however are very relevant.
The final expression for the effective Hamiltonian $\tilde{H}_{\rm eff}$ is
the following
\vspace{0.5cm}
\be \tilde{H}_{\rm eff}=
\left( \begin{array}{cccc}
      \vspace{0.5cm}
-p \cdot J^{\nu_e}_{\rm eff} + {1\over 2} \Delta m^2_{12} \sin^2\vt &
{1\over 4} \Delta m^2_{12} \sin 2\vt  &
E \m_{ee} \Om^\ast &
E \m_{e\mu}\Om^\ast \\
      \vspace{0.5cm}
{1\over 4} \Delta m^2_{12}\sin 2\vt  &
-p \cdot J^{\nu_\mu}_{\rm eff}+ {1\over 2} \Delta m^2_{12} \cos^2\vt &
E \m_{\m e} \Om^\ast &
E \mu_{\mu \mu} \Om^\ast \\
      \vspace{0.5cm}
E \m_{ee} \Om &
E \m_{e \mu}\Om &
0 &
0 \\
E \m_{\mu e} \Om &
E \m_{\m \mu}\Om &
0 &
{1\over 2} \Delta m^2_{12}
\end{array} \right )
\vspace{0.5cm}
\label{gau}
\ee
where $\vt$ is the neutrino mixing angle, $J_{\rm eff}$ is given
in (\ref{jefa}), $E$ is the energy of the particle, and $\Om$
is defined in (\ref{defofOm}).
Note that, because $\nu_{e_R}$ and $\nu_{\mu_R}$ are sterile, {\it i.e.},
do not interact electroweakly with matter, they can be considered vacuum
mass eigenstates.

\begin{center}
\subsubsection*{3.3 Resonances}
\end{center}

Using (\ref{gau}) we can determine the AGN regions where resonant
transitions occur (a phenomenon reminiscent of
the MSW effect \cite{{lw},{ms}}). These
resonances are governed by the $ 2 \times 2$ submatrices of
(\ref{gau}) for each pair of states~\footnote{We will ignore multiple
resonances
which occur whenever three (or four) diagonal elements concide; such
resonances can occur only under very restricted circumstances.}.

The five possible resonances are obtained by equating the diagonal terms for
each submatrix and give rise to the following resonance conditions

\ba
{1\ov 2} \Delta m^2_{12} \cos 2\vt \; +
     \; \sqrt 2 \; ( p \cdot u ) \; G_F N_p \!&=&\! 0 \quad
       \!\!\! (\nu_{e_L}\!\!\! \rightarrow \!\!\nu_{\mu_L}) \label{resa} \\
{1\ov 2} \Delta m^2_{12} \cos^2 \vt \; +
   \; {( p \cdot u )\ov \sqrt 2} \; G_F N_n \; + \; p \cdot J_G \!&=&\! 0 \quad
     \!\!\!   (\nu_{\mu_L}\!\!\! \rightarrow \!\!\nu_{e_R})  \label{resb}   \\
-{1\ov 2} \Delta m^2_{12} \sin^2 \vt \; +
    \; {( p \cdot u )\ov \sqrt 2} \; G_F N_n \; + \; p \cdot J_G \!&=&\! 0
\quad
      \!\!\! (\nu_{\mu_L}\!\!\! \rightarrow \!\!\nu_{\mu_R}) \label{resc}   \\
{1\ov 2} \Delta m^2_{12} \cos^2 \vt \; +
    \; {( p \cdot u )\ov \sqrt 2} \; G_F (2 N_p - N_n) \; - \; p \cdot J_G \!
&=&\! 0 \quad
         \!\!\!   (\nu_{e_L} \!\!\!\rightarrow \!\!\nu_{\mu_R})  \label{resd}
\\
-{1\ov 2} \Delta m^2_{12} \sin^2 \vt \; +
    \; {( p \cdot u )\ov \sqrt 2} \; G_F (2 N_p - N_n) \;- \;  p \cdot J_G
\!&=&\! 0 \quad
        \!\!\!    (\nu_{e_L}  \!\!\!\rightarrow \!\!\nu_{e_R}) \label{rese}
\ea
As will be shown in the following section the
matter effects are very small compared to the gravitational effects
which makes $ J_G $ dominant whenever present (this makes irrelevant the
precise value of $N_n$ for the cases (\ref{resb}--\ref{rese}); only for
the MSW resonance \cite{lw,ms}, eq. (\ref{resa}) is the matter density
important).

In order to determine whether the above resonances induce significant
transition probabilities we consider the corresponding $ 2 \times
2 $ submatrices in detail. Each of these can, by a suitable subtraction
from the  diagonal term, be cast into the form
\be \left( \begin{array}{cc}
d & b \\
b & -d
\end{array} \right )    \label{mat}     \ee
where $d$ is related to the matter and gravitational part and
$b$ to the  magnetic field; their
explicit form is
\be
b = E \mu_\nu \Omega \qquad
d = - p \cdot J_{\rm eff} + \left(\Delta m^2 {\rm\ term} \right)
\ee
where $E, \m_\n$ and $\Omega$ are respectively the energy, magnetic moment
of  the neutrino and electromagnetic field, see (\ref{defofOm});
the ``$ \Delta m^2 $ term'' depends on the specific $ 2 \times 2 $ matrix
and can be easily obtained from (\ref{gau}); it is always smaller than $
| \Delta m_{12}^2 |$.

Resonances occur when $d$ vanishes, in which case the transition probability
is well described (for sufficiently slowly varying $d$ and $b$) by the
Landau-Zener approximation \cite{lz}
\be
P_{LZ} = \exp \left\{ -2 \pi^2 \, {\b^2 \ov \a } \right\}
\ee
where
\be
\beta =  b \;\vert_{\rm res}\;\;\;\; \a = \dot{d}\;\vert_{\rm res} \label{alph}
\ee
The condition for these resonances to induce an appreciable transition
probability is
\be   \b^2 \ge {\a\over 2\pi^2}. \label{cond} \ee

The presence of a magnetic field can also induce
coherent precession of the states which has been studied in sufficient
detail in \cite{okun}. 
The condition for this to occur is, for slowly varying $b$ and $d$ in
(\ref{mat}),
\be
 { \sqrt{b^2 + d^2 }  \over E } R \ge 2 \pi \label{preccond} 
\ee
where $R $ is
the magnitude of the region where the magnetic field is coherent.
In this case the transition probability is 
\be
 P_{\rm prec} \simeq { 1 \over 2} { b^2 \over b^2 + d^2 }\label{preccondi}  
\ee
which is significant when $ b > d $. So, while the presence of a
diagonal term requires a shorter distance $R$ to generate rapidly
varying phases, the very same effect decreases the transition
probability: neutrino interactions with matter
or gravitational effects effectively quench spin precession
\cite{moretti}.

\begin{center}
\subsubsection*{3.4 Estimates}
\end{center}

The orders of magnitude for the weak and gravitational currents are, from
(\ref{den}) and (\ref{jefb}),
\be
J_W  =  {G_F \; \r \over m_p} \sim 10^{-33} \; \r \;\; {\rm eV}^{-1} ,\;\;\;
J_G \sim R^{-1}
\ee
where $\r$ is in units of ${\rm eV}^4$.
According to (\ref{dena}) and (\ref{denb}) the order of $\r$ for typical cases
is $10^1-10^4 \;{\rm eV}^4$. 
Clearly from above and taking $ R \sim r_g $, 
the  gravitational current part is found to dominate
the weak current part for all relevant values of $ r_g  \; (10^{14}\; {\rm to}
\; 10^{20} {\rm eV^{-1}} )$.

The order of magnitude for the $m\Th$ term in the Hamiltonian
will be (considering that $\d_\m \sim R^{-1}$), 
\be
m\Th^b \sim m R^{-1}\sim m r_g^{-1}.
\ee
If we compare this term with $E\m_\n\Om$ for typical values ($E\sim 1$TeV,
$\Om\sim 10^4$ G, $r_g\sim 10^{18}$ eV${}^{-1}$),
for a very small magnetic moment $ \m \sim 10^{-19} \m_B$ and neutrino mass $m\sim 1$ eV, we get
\be
{E\m\Om\ov m\Th^b\tau_b} \sim 10^6.
\ee 
Therefore, in what follows  we will neglect the $\Th$ term.

The values for $ | \Delta m^2_{12} | $ corresponding to different
scenarios of neutrino oscillation are
\vspace{0.5 cm}
\begin{table}[h]
\begin{center}
\begin{tabular} {|l|c|c|} \hline
	&  $  \Delta m^2_{12} \cos^2 \vt \;\;(eV^2)$  & $ \Delta m^2_{12} \sin^2 \vt \;\; (eV^2)$ \\
\hline
Vacuum \cite{bo} &   $10^{-10}$ & $10^{-11}$	\\ \hline
Solar small angle \cite{bo} &  $10^{-6}$ &  $10^{-8} $ \\ \hline
Solar large angle \cite{bo} &  $10^{-6} $ & $10^{-6}$      \\ \hline
At. neutrinos \cite{bo} &  $10^{-3}$ &  $10^{-4}$      \\ \hline
LSND \cite{lsnd}     &  $1$ &  $10^{-3}$   \\ \hline
\end{tabular}
\caption{Approximate values for $\Delta m^2$ terms. }
\end{center}
\end{table}

The MSW resonances (\ref{resa}) correspond to
\be \Delta m^2_{12}  \sim 10^{-33} E \r \ee
which, since $\r\le10^4\; {eV}^4$ and $E\le10^{19}$ eV (in order to get an
appreciable
flux) corresponds to $\Delta m^2\le10^{-10} \ {\rm eV}^2$. Thus, aside from
these extremely small values, {\it the usual MSW scenario does not take
place.}

Since $p \sim E$, neglecting weak current part,
\be p \cdot J_{\rm eff}\sim{ E \ov r_g } \ee
For a typical value of energy $E \sim 1 \, $ TeV, beyond which
AGN neutrinos start dominating the sky,
and $r_g \sim 10^{20} - 10^{14}
 \; {\rm eV^{-1}}$ ($ M = 10^{10} M_\odot- 10^4 M_\odot$),\,\,
$p\cdot J_{\rm eff}\sim 10^{-8}-10^{-2} \, {\rm eV}^2$. The
orders of magnitude from the
previous table show that values of $ \Delta m^2 $ corresponding to the
solar and atmospheric entries would undergo resonances in the vicinity
of the black hole. Larger values of the mass difference (such as the
LSND entry in the above table) would resonate only at significantly
larger energies (above $ 100 \,{\rm TeV}$);
in order to have comparable resonance
for Atmospheric and LSND {\cite{lsnd}} values we need
even more energetic neutrinos
($\sim 10^2  \,$ TeV and $\sim 10^6 \,$ TeV
respectively). A point to note here is that all resonant transitions
do not occur simultaneously as can be seen from equations 
(\ref{resa}-\ref{rese}). At fixed conditions for instance, the transitions
$ (\nu_{\mu_L}\!\!\! \rightarrow \!\!\nu_{e_R}) $ and $ (\nu_{e_L}  \!\!\!\rightarrow \!\!\nu_{e_R})$ cannot both be realized together.

In order to determine whether the above resonances induce large
transition amplitudes we will need to estimate the magnitude of $ \Om$.
The models we consider all assume that the magnetic field is determined
by the ``equipartition'' condition \cite{zp}
that the external pressure is matched
by the magnetic one. The scale height for the pressure is $ \sim
r_g $ and so
we expect the
magnetic field to be uniform only
through distances of order $ r_g $; on larger scales it
will vary randomly. As we mentioned in section 1, the value of
the magnetic
field within these ``coherence patches'' is $ | \Om | \sim |  B  | \sim 10^4$ G.
Let us now define the scale height  $\L$ for the effective current terms
\be
\L=\left| {dl\ov {d(\ln p \cdot J_{\rm eff})}} \right|
\ee
which has units of the affine parameter $l$.
Then $ \alpha \sim \Delta m^2 / \Lambda $ and the condition
(\ref{cond}) becomes
\be
\mu_\nu \ge { 1 \over E B} \left| { \Delta
m^2 \over 2 \pi^2 \Lambda } \right|^{1/2} = \mu^{\rm res}_{\rm min} \label{cond2}
\ee
Resonant transitions will occur then if the relevant magnetic moment (or
transition magnetic moment)
satisfies this bound; this constraint depends, through $ \Lambda $,
on the type of metric and will be studied for some cases of interest in
section 4 below. Note that we have assumed that the magnetic field
remains constant over an interval of magnitude $ \sim \Lambda $.

Within this estimate the condition for coherent precessions (\ref{preccond}) becomes 
\be
( \mu B r_g )^2  + ( { \Delta m^2 r_g \ov E } + 1)^2 >  4 \pi^2 \label{dee}
\ee
and  the
transition probability (\ref{preccondi}) becomes 
large provided 
\be
\mu_\nu   \ge { 1 \ov {B r_g}} \left(  1 + { \Delta m^2 r_g \ov E  } \right ) = \mu^{\rm prec}_{\rm min}\label{preccondii}
\ee
which is the condition for coherent precession to generate a significant
number of helicity flips.

Comparing the different magnetic moment expressions obtained above we get
\be
{ \mu^{\rm prec }_{\rm min} \ov \mu^{\rm res }_{\rm min}} = 
\left( { \Delta m^2 r_g \ov E }\right)^{1/2} + 
\left( { \Delta m^2 r_g \ov E }\right)^{-1/2} \label{ratio1}
\ee
which implies $ \mu^{\rm prec }_{\rm min} 
\,{\raise-3pt\hbox{$\sim$}}\!\!\!\!\!{\raise2pt\hbox{$>$}}\,
2 \mu^{\rm res }_{\rm min}$
for any value of  $\Delta m^2 r_g/E$.

Though realistic calculations would require an accurate profile for the
magnetic field, the above results do provide useful order-of-magnitude
estimates of the relevant quantities.
We will use these expressions in section 4.2 to estimate the possibility
of transitions of each type.

\begin{center}
\subsection*{4 Neutrino oscillations in a Kerr space time.}
\end{center}

In this section we will apply the formalism developed to the interesting
case of a Kerr black hole.

\begin{center}
\subsubsection*{4.1. Hamiltonian for the Kerr metric.}
\end{center}

The gravitational field of the rotating black hole is given by the following
axially symmetric stationary Kerr metric \cite{ll}:
\ba
d s^2 & = &( 1 - { r_g r \ov \rho^2} ) d t^2 - { \rho^2 \ov \Delta} d r^2 -
\rho^2 d \theta^2 - ( r^2 + a^2 + {r_g r a^2 \ov  \rho^2} \sin^2{\theta})
\sin^2{\theta} d \phi^2  \nn	\\
&&+ { 2 r_g r a \ov \rho^2 }\sin^2{\theta} d \phi\; d t
\ea
where \be \Delta  = r^2 - r_g r + a^2 \;\;,\;\;
  \rho^2 = r^2 + a^2 \cos^2\theta \ee

This metric depends on two constant parameters $M$ and $a$, where $M$  is
the mass of the black hole and $a$ is related to the angular momentum by
the relation
 $ L_{BH} = ma $; $r_g$  is given by
$r_g=2M$ in geometrized units ($c=G=1$).
For $ a=0 $ the Kerr metric becomes the Schwarzschild metric in the
standard form.

We now consider the motion of a particle of
mass $m$ in a Kerr field \footnote{Included
only for completeness and to define the notation used later.}.
The Hamilton-Jacobi equation is fully separable in this case \cite{carter}.
Writing the action S as
\be S = - E t + L \phi + S_r (r) + S_\theta (\theta), \ee
where $E$ is the energy and $L$ the angular momentum of the particle.
We obtain two ordinary differential equations
\ba
&&{\left ( { d S_\theta \ov d \theta } \right )}^2 + {( a E \sin\theta - { L
\ov \sin\theta})}^2 + a^2 m^2 \cos^2\theta = K  \label{moma} \\
&&{ \left ( {d S_r \ov d r } \right )}^2 - { 1 \ov \Delta }{ ( r^2 E + a^2 E -
a L)}^2 + m^2 r^2 = -K  \label{momb}
\ea
where K is a new constant of motion. The four-momentum of the particle is
then
\be
p^a = \left( {\g_0}^{-1} E \;,\; {\g_1}^{-1} S'_r \;,\;
{\g_2}^{-1} S'_{\t}\; ,\; { L - g_{ 03} E / g_{00} \over k} \right)
\label{oldp}
\ee
where
\be
\g_0 = \sqrt{g_{00}} \qquad \g_i = \sqrt{-g_{ii}}  \qquad
k^2 = {g_{03}^2 \over g_{33}}- g_{33}
\ee
Our calculations involved a set of
tetrads in which the  momentum takes the  form $ p^a = E(1,0,0,1)$. The
corresponding tetrads are
\[ e^a_\mu = \\
 \left( \begin{array}{cccc}
\g_0 & 0 & 0 & \eta \\
0 & \g_1 c_{\a} & 0 & k s_{\a} \\
0 & - \g_1 s_{\b} s_{\a} & \g_2 c_{\b} & k s_{\b} c_{\a} \\
0 & - \g_1 c_{\b} s_{\a} & -\g_2 s_{\b} & k c_{\b} c_{\a}
 \end{array} \right ) \]
where
\be s_{\a}= \sin \a ,\;\; c_{\a} = \cos \a \ee
and similarly for $\b$; $ \a $ and $ \b$ are the polar and
azimuthal angles of $ p_a $ in (\ref{oldp}).
The angles fulfill the following conditions
\be
\tan \a = -{p^1 \over p^3} \;\;\;\; \tan \b =-{p^2 \over \sqrt{(p^1)^2 +
(p^3)^2}}.
\ee
Using the above relations we get an expression for the effective
Hamiltonian for Kerr metric (\ref{heff}). We use this to look at
resonances in the next section. The explicit expression opf the various
terms in the Hamiltonian is quite involved, we include a brief
description of such terms in Appendix D; here we present the result of
numerical analysis of the expressions.

As a limiting case of the previous analysis, we consider the limit $ a 
\rightarrow 0 $, corresponding to a Schwarzschild black-hole. The
neutrino geodesics lie on a plane which we take as the $ \theta = \pi/2
$ plane, we then obtain $ p_a \bar J^a_G = 0 $ and the effective
Hamiltonian reduces to 
\be
\tilde{H}_{\rm eff} = {1\over2} m^2 - { E \bar J_W^{ a = 0 } \over \sqrt{ 1 - r_g/r } }
\left( \begin{array}{cc} 0 & 0 \\ 0 & 1 \end{array} \right) +
\mu \left( \begin{array}{cc}
0 & \Omega^* \\ \Omega & 0 
\end{array} \right)
\ee 
where we assumed the absence of matter currents and where
$ m $ and $ \mu $ denote the mass and magnetic moment matrices
respectively. Thus, except for a factor $ 1 / \sqrt{ 1 - r_g/r } $ which
is important only for $ r \sim r_g $ the above expressions reduce to the
flat-space situation. In particular there is no gravitational
contribution to the diagonal elements. One can also easily determine the
off-diagonal terms $ \Theta^b $ in (\ref{jefc}), which give a
contribution $ m L \tau_2 / ( 2 E r^2 ) $ to $ \tilde{H}_{\rm eff} $ ($ \tau_2 $
denotes the usual Pauli matrix). As mentioned previously these terms are
negligible for all cases of interest due to the factor of $ m / E $.

\begin{center}
\subsubsection*{4.2 Resonances for the Kerr metric}
\end{center}

 We need to determine whether resonances are possible  in the vicinity
of a Kerr black hole. For a rotating black hole the spin flavor
transition is due to the transference of orbital angular momentum to
spin angular momentum and also due to the transference of the black hole's
angular momentum to the neutrino spin.
To determine possible resonant behavior, we will find the regions  ($r,\t$)
where each of equations (\ref{resa}) to (\ref{rese}) are satisfied as a
function of the parameters $ E, L, K, a, r_g $. We will do this for different
values of $ \Delta m^2_{12}$ and  mixing angle.
We choose  normalized parameters
\be \j ={ L \over E r_g } , \qquad \k={ K \over (E r_g)^2} 
\ee for a neutrino obeying the equations (\ref{moma}) and
(\ref{momb}).
It is easy to see that $p \cdot J_{\rm eff}$ can be written (neglecting
the matter terms) in terms of the energy and Schwarzschild radius
in the form
\be
\vert p\cdot J_{\rm eff} \vert = Er_g^{-1} f(r/r_g,\t,\j,\k, a/r_g).\label{resca}
\ee

In Fig. 1 we show $ \ln | f | $ as a function of $(r,\t)$, for an allowed
value of the pair $( \j,\k)$, where an  angular momentum
$(a = 0.4 r_g)$ has been chosen for the black hole.\footnote{A realistic
analysis of accretion onto black holes must account for the fact that the
central hole is quite probably a Kerr black hole with angular momentum
parameter $ a $ only slightly less than the gravitational mass \cite{shap}.}
We have plotted the magnitude of $ p\cdot J_{\rm eff}$ in an effort to
give a clear picture of how resonances take place in this particular
metric. It is to be borne in mind however that all resonance transitions
do not occur simultaneously.
Figure 2 presents the $f$ contour plots under the same
conditions as in Fig. 1.

\setbox2=\vbox to 160 pt {\epsfysize=6 truein\epsfbox[0 -200 612 592]
{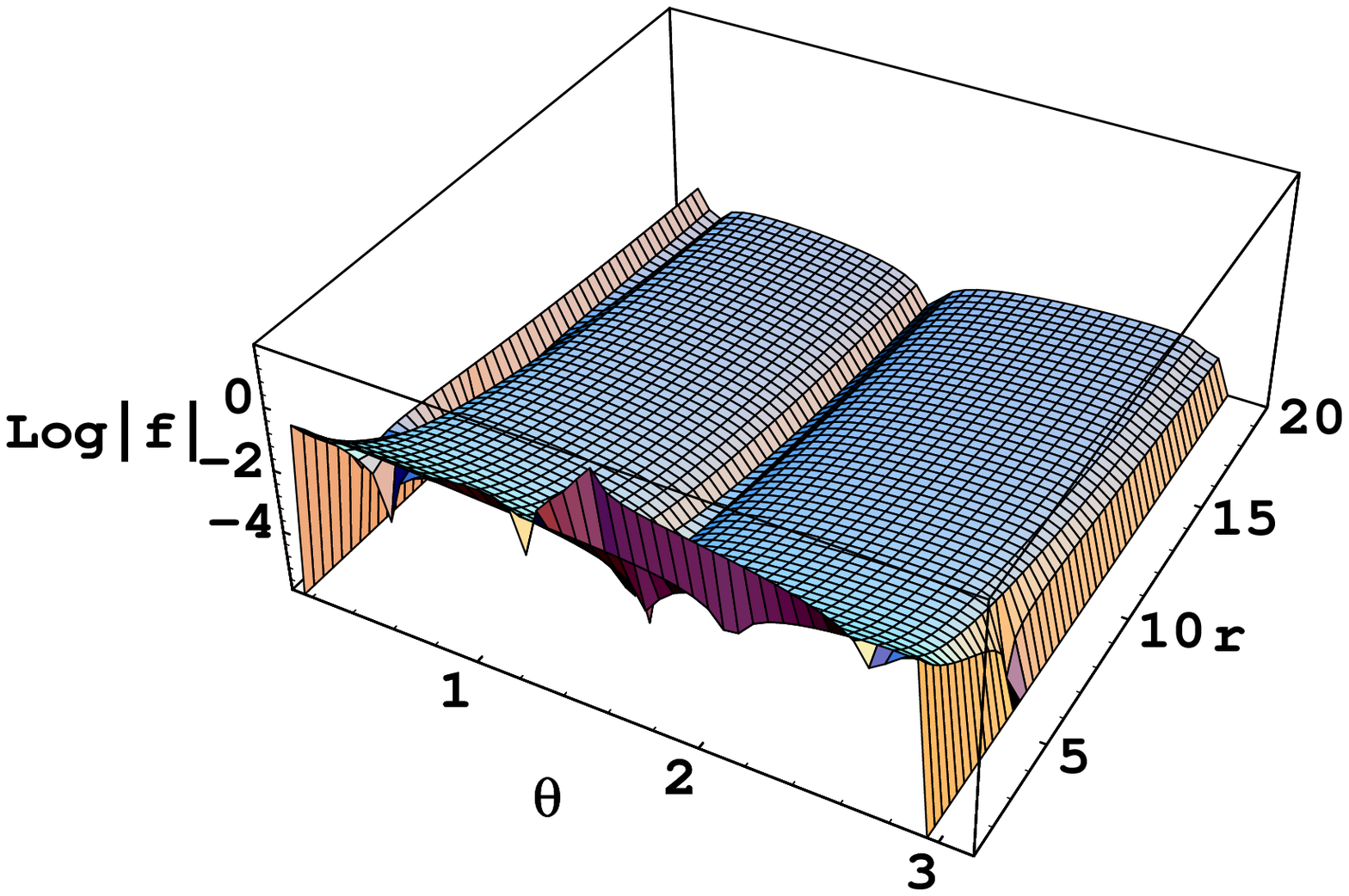}} \centerline{\box2}

\thecaption{Fig.1.}{Plot of $ \ln |f|$, see (\ref{resca}), for $\j=0.4, 
\; \k=0.15$, $a/r_g=0.4$ and $M=10^8 M_\odot $.}

\bigskip\bigskip\bigskip

\eject

\setbox2=\vbox to 160 pt {\epsfysize=4 truein\epsfbox[0 -200 612 592]{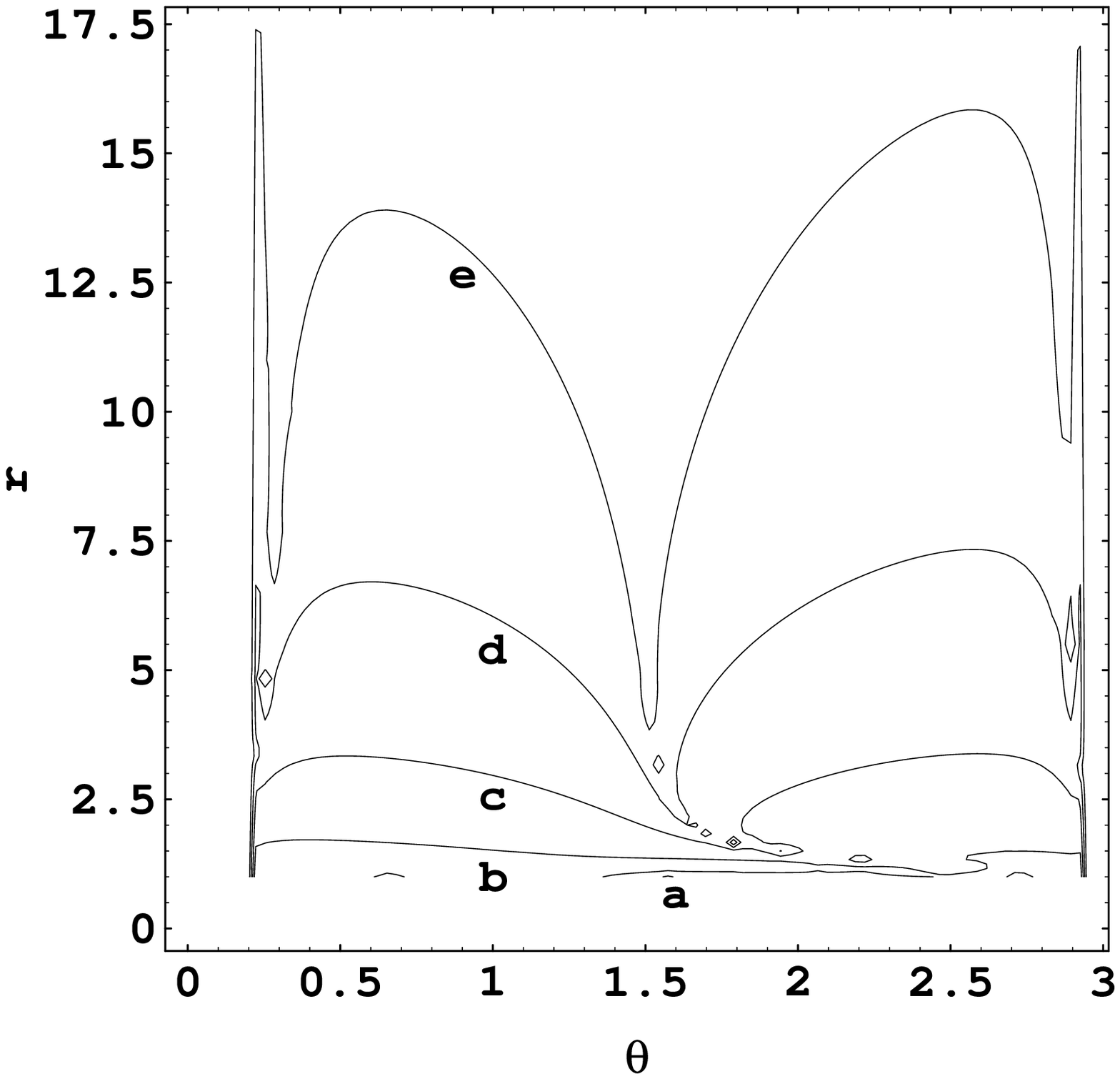}}
\centerline{\box2}

\thecaption{Fig.2.}{Contour-plot of $f$, see (\ref{resca}), for $\j=0.4,\; 
\k=0.15$,  $a/r_g=0.4$ for the case  $M=10^8 M_\odot $.
The contours a,b,c,d,e correspond respectively to
$f(a)=\pm 1,\,f(b)=\pm 10^{-1},\,f(c)=\pm 10^{-2},\,f(d)=\pm 10^{-3},\,
f(e)=\pm 10^{-4}$, 
where positive values take place for  $0\le \theta \le \pi /2$, and
the negatives ones for $\pi /2 \le \theta \le \pi$.}

\bigskip\bigskip

\setbox2=\vbox to 160 pt{\epsfysize=6 truein\epsfbox[0 -200 612 592]{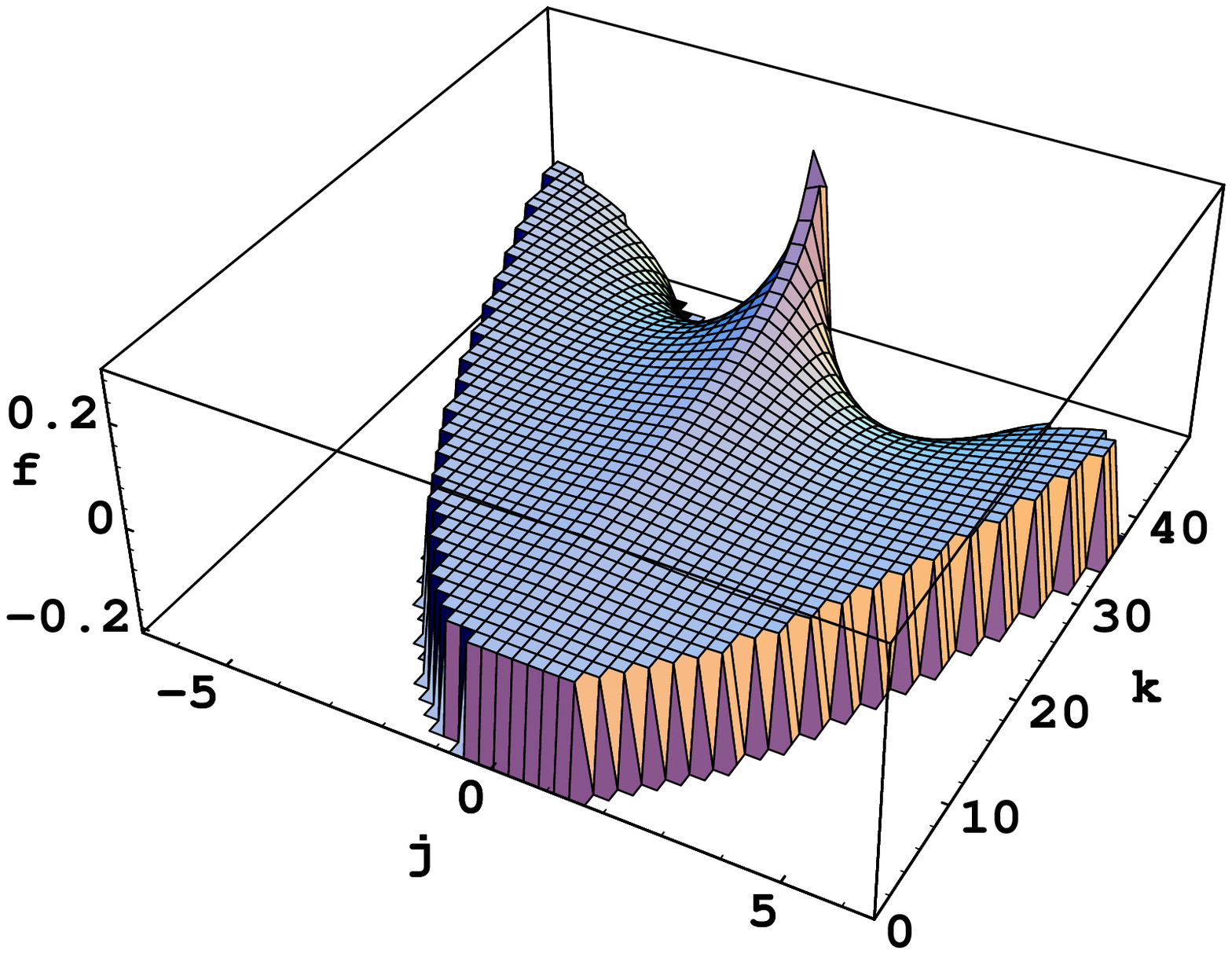}}
\centerline{\box2}

\thecaption{Fig.3.}{Plot of $f$, see (\ref{resca}), for $r=6r_g,\; \t=\pi/2$ , 
$a/r_g=0.4 $ and $M=10^8 M_\odot $.}

\bigskip\bigskip

Figure 3 illustrates the dependence of the function $f$
for a specific value of  $(r,\t)$, on allowed values
of $(\j,\k)$ and taking $a = 0.4 r_g$.

Resonances occur provided $f$ is comparable to $\pm \Delta
m^2_{12} \cos^2 \vt\,r_g/E$ or $\pm \Delta m^2_{12} \sin^2 \vt\,r_g/E$
as can be seen from equations (\ref{resb}) to (\ref{rese}).
As an example we can analyze the case of energy $E \sim 1$ TeV,
 $r_g\sim 10^{18} {\rm eV}^{-1}$ and consider
the Solar large angle solution,  $\Delta m^2_{12} \sim 10^{-6} $eV$^2$.
Comparing
these with $f$ from figure 2, we find  that it corresponds to contour ``a''.
We  conclude then that resonances occur in the vicinity of the AGN for
this choice of energies.
In the same way it can be seen that resonances are present for all
the given values of $ \Delta m^2 $ in Table 1,
all relevant black-hole masses and angular momenta, and all neutrino
energies above $ 1 {\rm TeV } $.

It is also of interest to determine the energy dependence of the
resonance conditions, which is presented in figures 4 and 5.

\setbox2=\vbox to 160 pt{\epsfysize=5.5 truein\epsfbox[0 -200 612 592]{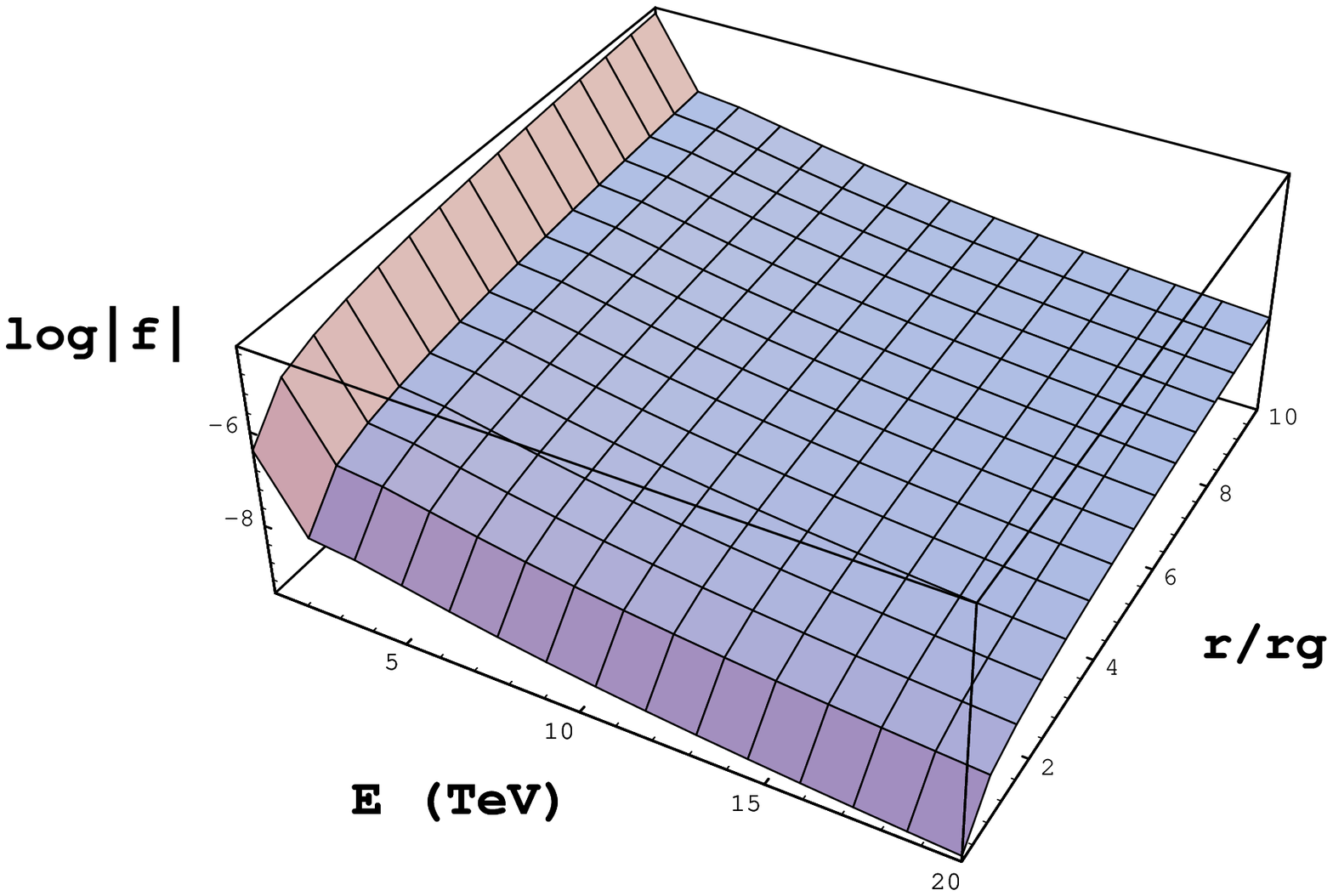}}
\centerline{\box2}

\thecaption{Fig.4.}{Plot of $ \log|f| $, see (\ref{resca}), as a function of 
$r$ and $E$ for the case $\theta=\pi/4$,
$ \j=0.4, \; \k=0.15$, $ a/r_g=0.4$ and $ M = 10^8 M_\odot$.}

\setbox2=\vbox to 160 pt{\epsfysize=5.5 truein\epsfbox[0 -200 612 592]{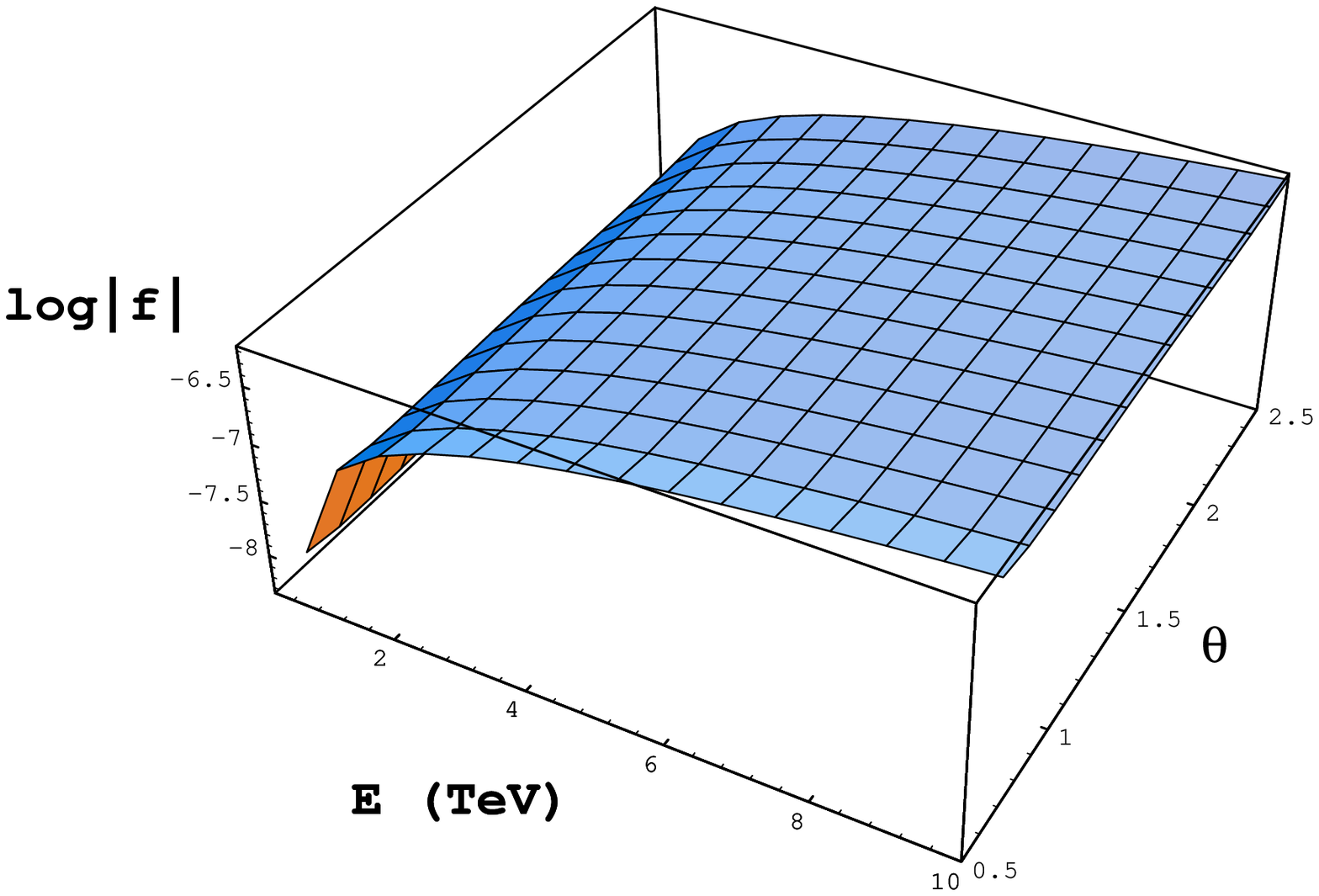}}
\centerline{\box2}

\thecaption{Fig. 5.}{Plot of $ \log|f| $, see (\ref{resca}),
 as a function of $\theta$ and $E$ for the case $r=6 r_g$,
$ \j = 0.4, \; \k = 0.15$ and $ a/r_g = 0.4$ and $ M = 10^8 M_\odot$.}

\bigskip
For fixed values of $j,\, k,\, r$ and $ \theta $, the function $f$
depends linearly on the energy $E$.

\medskip

In order to determine the values of the magnetic moment which will allow
for these resonances to induce large transition probabilities, we
defined the quantity $ \m^{\rm res}_{\rm min}$ in equation (\ref{cond2}).
This is plotted in Figs. 6  and 7 for some representative values of
$ \Delta m^2 $ and $E$ and for a magnetic field of $10^4 $G.

\eject

\setbox2=\vbox to 160 pt{\epsfysize=5 truein\epsfbox[0 -200 612 592]{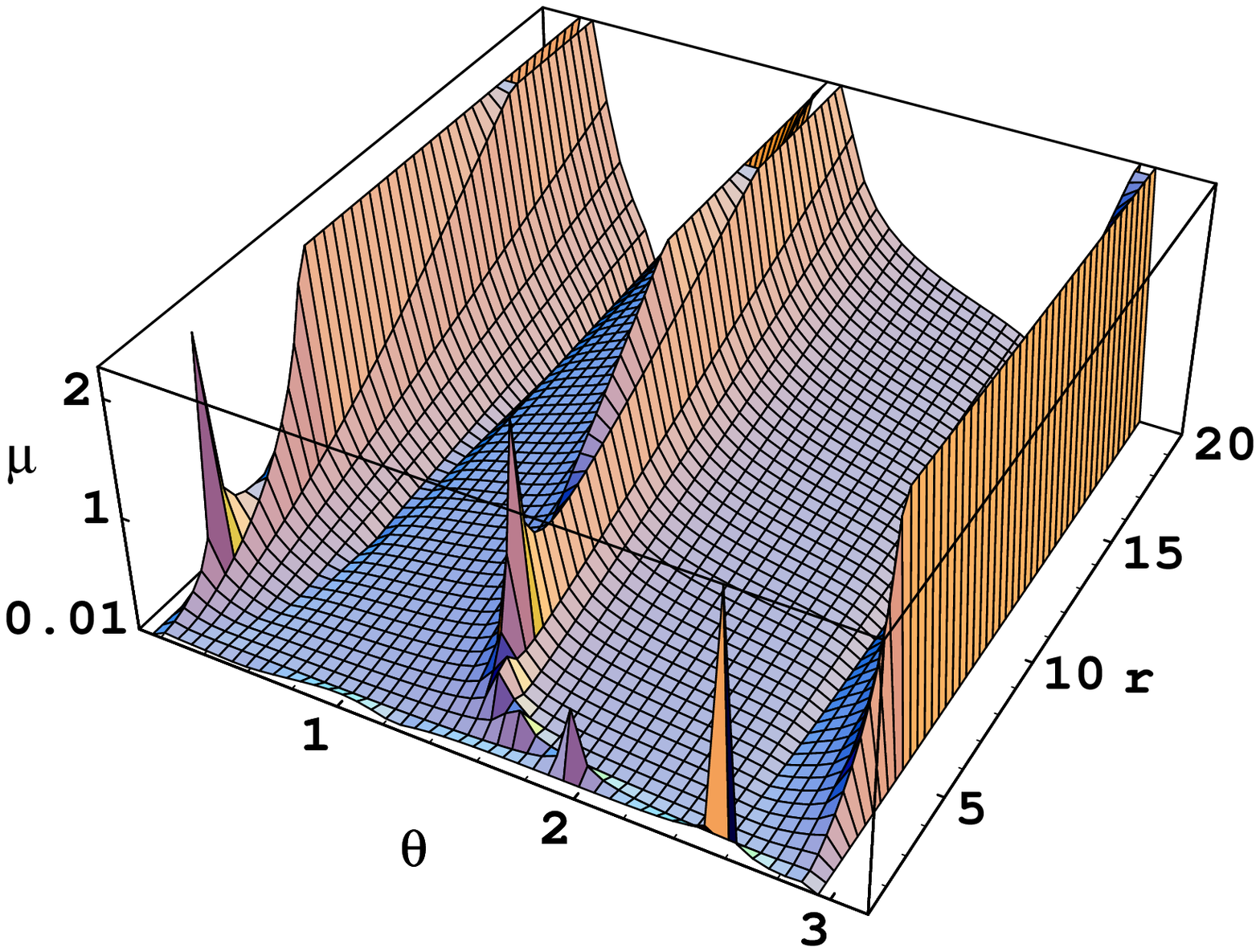}}
\centerline{\box2}

\thecaption{Fig. 6.}{Plot of $10^{13} \m^{\rm res}_{\rm min}$, see 
(\ref{cond2}),
for the case $ \Delta m^2 = 10^{-6}$ eV${}^2$ ,
\,$E=1$TeV,\, $ \j = 0.4, \; \k = 0.15$,   $ a / r_g = 0.4$ and
 $ M = 10^8 M_\odot$.}

\bigskip\bigskip

The resonances which we described above will induce large transitions
provided $ \m_\n $ is larger than $ \mu^{\rm res}_{\rm min} $; for the example
considered this corresponds to $ \sim 10^{-13} - 10^{-14} \mu_B $ 
(where $ \mu_B $ denotes the Bohr magneton). 
The above requirements
lie comfortably inside the direct experimental bounds
($ \mu_\nu \le 10^{-10} \mu_B $)
\cite{PRD} as well as the astrophysical limits ($ \mu_\nu \le 10^{-11}
\mu_B$) \cite{PRD}
 and such values for the
magnetic moment are consistent with a wide variety of models \cite{mb}.
In view of
this the above resonances will induce significant transition
probabilities whenever the resonance conditions (\ref{resa}-\ref{rese})
are satisfied.

\setbox2=\vbox to 160 pt{\epsfysize=5 truein\epsfbox[0 -200 612 592]{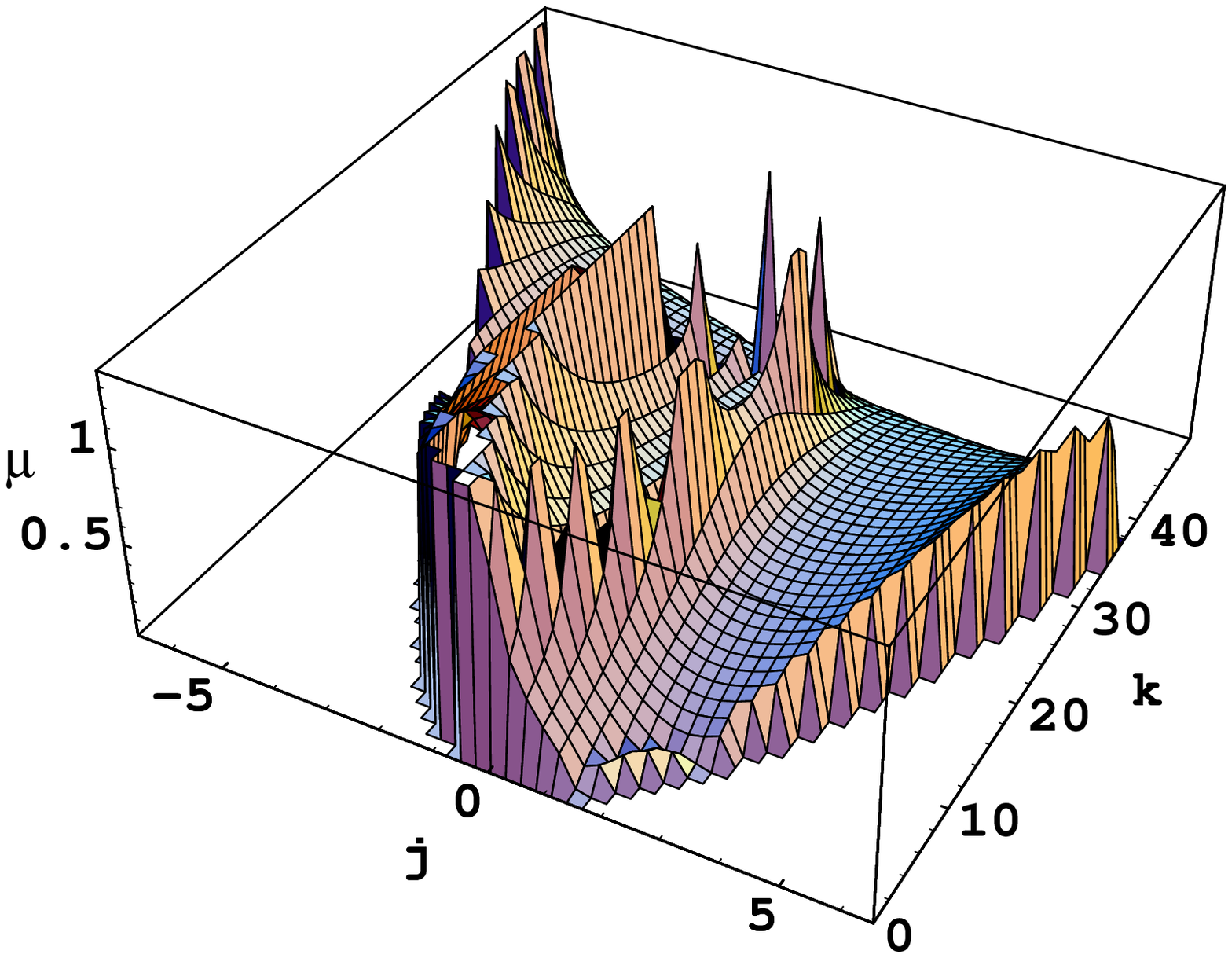}}
\centerline{\box2}

\thecaption{Fig. 7.}{Plot of $10^{13} \m^{\rm res}_{\rm min} $, see (\ref{cond2}), for the case $ \Delta m^2 = 10^{-6}$ eV${}^2$, \,$E=1$TeV,$\;
r = 6 r_g, \; \t = \pi /2$, $ a / r_g = 0.4$ and  $ M = 10^8 M_\odot$ .}

\bigskip\bigskip

The minimum of $ {\mu^{\rm prec}_{\rm min}} /{ \mu^{\rm res}_{\rm min}}$ 
(\ref{ratio1}) (note that equation (\ref{ratio1}) involves only an order of magnitude
calculation where $|f| \sim 1$, (\ref{resca}))
corresponds to $ \Delta m^2 r_g/E \sim 1 $,  and then $\mu^{\rm prec}_{\rm min}$
is comparable to $\mu^{\rm res}_{\rm min}$; for $ E =  1 $TeV 
and $r_g \sim 10^{18}$eV${}^{-1}$ this corresponds to 
 $ \Delta m^2\sim 10^{-6} {\rm eV^2 }$ which are the parameters used in plots 6 and 7.
Explicit calculations show that
$ \mu^{\rm prec}_{\rm min} > \mu^{\rm res}_{\rm min}\;$ whenever $\; |f| >
1$, which at resonance corresponds to $\Delta m^2 r_g / E > 1 $. 
The transition efficiency is greatest for the minimum value of the magnetic
moment and corresponds to the spin flip in vacuum.

In order to estimate the effects of these resonances on the neutrino
fluxes on Earth we note that,
neutrinos are expected to be generated mainly through the $ \pi
\rightarrow \mu \rightarrow e $ decay
chain, implying that we can expect twice as many muon-type
neutrinos as electron-type neutrinos. A  negligible number of tau
neutrinos are produced in an AGN environment. Even though the above
calculations were done for the case of two species, we expect the same
results for the case of three species: resonances will occur for all
possible transitions. Such resonances will tend to ``spill over'' the
neutrinos from the highly populated states into the less populated ones
evening out the neutrino numbers in all states. It follows that, with
these simplistic assumptions, the muon-neutrino flux will be decreased
by a factor of $ 1/4 $, the electron neutrino flux will decrease by a
factor of $ 1/2 $ but we can expect as many tau-neutrinos.

\begin{center}
\subsection*{5. Conclusions }
\end{center}
The evolution of neutrinos in an AGN environment is
susceptible to gravity-induced resonances. Such resonances would induce
various spin and spin-flavor transitions whenever equations
(\ref{resb}--\ref{rese}) are satisfied. We have shown that such
conditions would be expected to be satisifed for the currently accepted
AGN models and for various neutrino parameters currently used in the
literature. 

The lack of precise modeling of the AGN magnetic field (eg. is the field
dispersed or is it concentrated in flux tubes), and lacking a better
understanding of the parameters in the neutrino system it is impossible
to determine unambiguously whether such resonances would in fact occur
in a given AGN.

Using the estimates for the magnetic fields, and
approximating the gravitational field by the Kerr metric, the various
values for $ \Delta m^2 $ we use do lead to resonances of the type 
(\ref{resb}--\ref{rese}). The corresponding transition probabilities
will be large for a sufficiently large magnetic moment. Each resonance
will occur at a different place along the neutrino trajectory;
plots 1, 2 and 3 show that essentially all neutrinos produced
near the AGN core will traverse regions where all four resonance
conditions are satisfied (for the neutrino parameters under
consideration). Such resonances induce large transition probabilities
for a wide variety of neutrino masses and energies (provided the
neutrino magnetic moment is sufficiently large).

We do not expect resonant transitions to be 100\%\ effective, whence the
neutrinos will tend to populate all available states. Lacking detailed knowledge
of the AGN environment, but relying on the fact that all resonances are
viable, we can only estimate that all such states would
end up by being equally populated. This, of course, depends on the
masses of the neutrino states. For very small 
values of $ \Delta m^2 $, $ \Delta m^2 < 10^{-10}{\rm eV}^2 $,
(\ref{resa}) will be satisfed and MSW
resonances would occur; the other posibilities (\ref{resb}--\ref{rese})
can be realized, but now farther from the AGN core where the estimate of
the gravitational field is less reliable and where matter effects can no
longer be ignored. We will not consider this (very complex) possibility 
further in this paper

It is interesting to note that for $ \Delta m^2  > 10^{-10}{\rm eV}^2 $,
the resonances described in this paper are not related to
the MSW resonant flavor oscillations (which are suppressed in this
environment due to the low densities present
unless $ \Delta m^2 < 10^{-10}{\rm eV}^2 $, see sect. 3.2), but to the interplay of
the gravitational and electromagnetic interactions with the mass terms.
Also it must be noted that
all resonance transitions do not occur simultaneously
as can be seen studying equations (\ref{resa}-\ref{rese}).

Noting that the AGN neutrinos travel through cosmological distances
before being observed on Earth and taking the
matter effects to be negligible on the way to the
earth one might argue that the vacuum
oscillations generated by the non-zero mass matrix would produce
identical effects to the ones considered here. For a distance of
approximately  100 Mpc (a sensible distance to AGN), vacuum oscillations 
occur for 
$ \Delta m^2 < 10^{-19}  {\rm{eV}}^2 $ for 1 TeV energy neutrinos
which is restrictively small.
The effects of a cosmic magnetic field 
(the typical value of galactic \cite{galmf} and galactic cluster 
\cite{clumf} magnetic fields 
is of the order of $10^{-6}$G; the
intercluster magnetic field is $ \lesim 10^{-9}$G \cite{cosmf})
can, of course, induce rapidly varying phases in the neutrino states,
but the transition probability (\ref{preccondi}) is nonetheless small.

In our calculations we have chosen a rapidly rotating black hole ($ a
= 0.4 r_g$). This assumption is based on the conclusions of reference
\cite{shap}. For lower values of $a$ the conditions under which the
resonances discussed in ths paper can occur become more stringent. For
the extreme limit of a Schwarzschild back hole, transitions are expected
only in the immediate vicinity of the horizon.

In conclusion we can say that gravitational induced resonance transitions are
the most important processes which can generate effective spin flavor conversion
in Active Galactic Nuclei.
The precise expressions for the flux to be detected at Earth-bound
neutrino telescopes depends, of course, on the details of the AGN
model. We have shown, nonetheless that for a very wide range of neutrino
masses and even with very small magnetic moments, neutrinos will undergo
resonant transitions. In their trip through the AGN environment the
neutrinos can experience several such transitions which tends to
even-out the population of all neutrino states. On one hand this
decreases the expected flux of electron and muon neutrinos, on the other
hand it dramatically increases the number of tau neutrinos of energies
of 1 TeV and above. This provides strong motivation to search for
such tau neutrinos by the existing neutrino telescopes, and the  recently
proposed 1 KM3 detector \cite{pak}. Even if the magnetic moments are $
\sim 10^{-14} \mu_B $ we can expect an observable number of, for example, $
10 $ TeV $ \tau $ neutrino events; the precise number of such events is
determined by the flux of electron neutrinos at that energy.

\par\vfill\eject

\newpage

\appendix

\begin{center}
\section*{Appendix A}
\end{center}
\setcounter{equation}{0}
\def\theequation{A.\arabic{equation}}

In this appendix we give a derivation of equation (\ref{pnu}),
Consider a geodesic $ \bar  x_\mu ( l ) $ for which the classical
momentum is $ \dot { \bar x^\mu } = p^\mu $ and solves the
Hamilton-Jacobi equation $ g_{\m \n} p^{\m} p^{\n} = 0 $. Taking an $l$
derivative we then obtain
\be
2 \dot p_\mu =
{ \partial g_{ \alpha \beta} \over \partial x^\mu } p^\alpha p^\beta
\label{apai}
\ee
Expanding now $x$ to first order around the geodesic,
$ x^{\m} = \bar{x}^{\m}+\n^{\m}_{A} \xi^{A} $ (recall that
this expression describes geodesics to $ O ( \xi^2 ) $),
and substituting in the Hamilton-Jacobi equation yields ,
\be
 2 \dot p_\mu \dot \xi^\mu +
{ \partial g_{ \alpha \beta} \over \partial x^\mu } p^\alpha p^\beta
\xi^\mu = 0 ,
\label{apaii}
\ee
where $ \xi^\mu = \nu^\mu_A \xi^A $.
From (\ref{apai}) and (\ref{apaii}) it immediately follows that
\be
0 = p_{\m} \dot{\xi}^{\m}+ \dot{p}_{\m} \xi^{\m} =
{d\over dl}(p_{\m} \xi^{\m}) .
\ee
Using the fact that the $ \xi^A $ are independent of $l$ it follows that
$ p_\mu \nu^\mu_A = c_A = $constant as was to be shown.

\begin{center}
\section*{Appendix B}
\end{center}
\setcounter{equation}{0}
\def\theequation{B.\arabic{equation}}

In this appendix we show how to obtain a Schr\"{o}dinger-like equation from
equation (\ref{red1}).
We note that
\be
i \dot{U}_0+i \dot{U}_{1 \over 2} =
{\cO} (U_0 + U_{1 \over 2})
-{ im \ov 2} \Us^A V_{1A} - { m \ov 2} \bar{\Vc}_0 U_0.       \label{prel}
\ee
But, from (\ref{expa})
\ba
i \dot{U}_0+i \dot{U}_{1 \over 2} &=& i \dot{\chi} + O(1/R^2)   \nn  \\
{ im \ov 2} \Us^A V_{1A}  &=& { im \ov 2} \Us^A \d_A \chi + O(1/R^2)
\ea
and therefore by reordering terms in(\ref{prel})
\be
i(\d_l+ {m \ov 2} \Us^A \d_A )\chi = \tilde{\cO}\chi+ O(1/R^2) \label{bo}
\ee
where $\tilde{\cO}$ is given by
\be
\tilde{\cO } = {\cO}- {m \ov 2} \bar{\Vc}_0, \label{ocal}
\ee
The second term in  (\ref{bo}) describes the spreading of the wave
packets and is not relevant for our discussion. We therefore ignore such
terms. The final equation therefore reads,
\be i \dot\chi = {\tilde{\cO }} \chi       \ee

\begin{center}
\section*{Appendix C}
\end{center}
\setcounter{equation}{0}
\def\theequation{C.\arabic{equation}}

In this appendix we evaluate the effective Hamiltonian, given by equation
(\ref{eff}), $
H_{\rm eff}  P_+ =  P_+ \left( {\cO}- m \bar{\Vc}_0 /2 \right) P_+
$
where $ {\cO } , \; \bar{\Vc}_0 $ are defined in (\ref{red2})
and (\ref{pote}). The explicit expressions are
\ba
{\cO } &=& i \a + {m^2\over 2} - ip^a \bar{\g}_{abc} \s^{bc} - p_a (\bar{J}^a
-{1\over 2}\e^{abcd} \bar{\g}_{bcd}) P_L      \nn 	\\
\bar{\Vc}_0 &=& (-{i\over 2} \g^{ba}_{\;\;\;b}+{1\over 2}\bar{ J}^a) \g_a -
({1\over 4}\e^{abcd} \bar{\g}_{bcd} +{1\over 2}\bar{ J}^a) \g_a \g_5
\ea
where
\be
\bar{\g}_{abc} = \bar{e}_{ai;j}\bar{e}_b^i \bar{e}_c^j
; \qquad
\s^{ab} = {1 \ov 4} [ \g^a , \g^b].
\ee
Using the explicit form of $ P_+ $ and working in the chiral
representation for the gamma matrices we obtain
\ba
P_+ \g^a P_+ &=& i \e^{abcd} \; \tau_b \otimes \tau^1\; {\pp_c p_d \over \ppdo}
P_+	\nn	\\
P_+ \g^a \g_5 P_+ &=& ( -\eta^{ab}+{p^a \ppb + \ppa p^b -p^ap^b \over \ppdo})
\; \tau_b \otimes \tau^1\; P_+	\nn	\\
P_+ \s^{ab} P_+ &=& {1\over 2 \ppdo}(\ppa p^b - \ppb p^a - i \e^{abcd} p_c
\pp_d
\g_5)P_+
\ea
when $ p^a = E ( 1, 0 , 0 , 1 ) $ and $ \ppa = E ( 1 , 0 , 0 , 0 )$.
Using these relations we obtain
\be
H_{\rm eff} = i \dot{\alpha} + { 1 \ov 2} m^2 - p \cdot J_{\rm eff} P_L +
m \Th^b \tau_b.
\ee
where $J_{\rm eff}$ is given by (\ref{jefb}) and $ \Th^b$ by (\ref{jefc}).

\begin{center}
\section*{Appendix D}
\end{center}
\setcounter{equation}{0}
\def\theequation{D.\arabic{equation}}

In this appendix we show how to calculate the $p\cdot J$ term in the 
Hamiltonian; the metric is given in section 4.1.
We will use the following notation
\be
\g_0 = \sqrt{g_{00}} \qquad \g_i = \sqrt{-g_{ii}}  \qquad
k^2 = {g_{03}^2 \over g_{33}}- g_{33} \qquad \eta = { g_{03} \over
\gamma_0}
\ee
along with the tetrads
\[ e^a_\mu = \\
 \left( \begin{array}{cccc}
\g_0 & 0 & 0 & \eta \\
0 & \g_1 c_{\a} & 0 & k s_{\a} \\
0 & - \g_1 s_{\b} s_{\a} & \g_2 c_{\b} & k s_{\b} c_{\a} \\
0 & - \g_1 c_{\b} s_{\a} & -\g_2 s_{\b} & k c_{\b} c_{\a}
 \end{array} \right ) \]
\be e_{a\mu} = \et_{ab} e^b_\mu.\ee
and
\[ e_a^\mu = \\
 \left( \begin{array}{cccc}
\g_0^{-1} & 0 & 0 & 0 \\
-k \Xi c_{\a} s_{\b} & \g_1^{-1} c_{\b} & -\g_2^{-1} s_{\a} s_{\b} &
-k^{-1} c_{\a} s_{\b} \\
-k \Xi s_{\a} & 0  & \g_2^{-1} c_{\a}  &  -k^{-1} s_{\a} \\
k \Xi c_{\a} c_{\b} & \g_1^{-1} s_{\b} & \g_2^{-1} s_{\a} c_{\b} &
k^{-1} c_{\a} c_{\b}
 \end{array} \right ) \]
where
\be \Xi= g_{03} (g_{00} g_{33} - g_{03}^2)^{-1} \ee
With these expressions we construct $\lambda_{abc}$ according to
\be
\lambda_{abc} = ( e_{a\m ,\n} - e_{a\n,\m}) e^{\m}_b e^{\n}_c.
\ee
Explicitly
\be
p\cdot J = p\cdot (J_W + J_G),
\ee
where the currents are given in (\ref{jefb}) and (\ref{den}),
being the weak current part negligible compared to the gravitational part
(as shown in subsection 3.4), and
\be
p\cdot J_G \simeq {1\over 2}E(\la_{132}+\la_{213}+3\la_{321}+3\la_{021} +\la_{
102}+\la_{210})
\ee
where, for example,
\be
\la_{132}=[{s_\a s_\b\ov {k^2 \g_1}}(k-c_\a c_\b\g^2_1)\d_1+
{c_\b\ov {k^2 \g_2}}(k s^2_\a + \g^2_2 c^2_\a)\d_2 
+{s_\a c_\a c_\b\ov k^2}(1+ k c_\b )\d_3 \; ]  k s_\b.
\ee
The angles fulfill
\be
\tan \a = -{p^1 \over p^3} \;\;\;\; \tan \b =-{p^2 \over \sqrt{(p^1)^2 +
(p^3)^2}}.
\ee
In the expression for $\lambda_{abc}$, $\d_i$ with $i=1,2,3$ represents
$\d_r, \d_\theta$ and
$\d_\phi$ respectively.
Similar expressions hold for the other values of  $\lambda_{abc}$, which
we will not list.

\end{document}